\documentstyle[preprint,prd,aps,12pt]{revtex}

\preprint{
$
\begin{array}{r}
\text{LAVAL-PHY-96-17} \\ 
\text{ALBERTA-THY-39-96} \\ 
\end{array}
$
}
\tighten

\begin{document}
\author{B. Dion, L. Marleau, G. Simon}
\address{D\'epartement de Physique, Universit\'e Laval\\
Qu\'ebec QC Canada, G1K 7P4}
\author{M. de Montigny}
\address{Facult\'e Saint-Jean, University of Alberta\\
Edmonton AB Canada, T6C 4G9}
\title{Scalar Leptoquark Pair Production at the CERN LHC:\\
Signal and Backgrounds}
\date{1996}
\maketitle
\draft

\begin{abstract}
We present the results of an analysis for the pair production of scalar
leptoquarks at the CERN Large Hadron Collider (LHC) with $\sqrt{s}=14$ TeV
and ${\cal L}=10$ fb$^{-1}$ which includes the dominant sources of Standard
Model background associated to this process: $t{\bar{t}}$, $ZZ$, $WZ$ and $%
Z^{*}jj$ production. The $t{\bar{t}}$ process provides the main source of
background. We consider leptoquarks introduced in the framework of a
superstring-inspired $E_6$ model. The leptoquark production is found to be
dominant in all regions of parameter space for leptoquark masses below 750
GeV. We establish the discovery reach of the leptoquarks at $750$ GeV ($1$
TeV) for a branching ratio of $B(LQ\rightarrow eq)=0.5$ ($B=1$).
\end{abstract}


\section{\bf Introduction}

H1 \cite{H1} and ZEUS \cite{ZEUS} experiments have recently reported an
excess of deep inelastic neutral current events in the range $Q^{2}\geq
15000 $ GeV$^{2}$. This has prompted several theoretical and
phenomenological analyses \cite{heraq2} seeking a proper interpretation. One
such interpretation for these events suggests single scalar leptoquark
production in the $e^{+}q$ or $e^{+}{\bar{q}}$ channels. Although the
statistics for these high-$Q^{2}$ events remain quite low for now (12
events) and no confirmation can be drawn until further measurements are
performed, it is nonetheless interesting to look at the discovery
possibilities of scalar leptoquarks at existing or future hadron colliders.

Leptoquarks are known to occur in various extensions of the SM, such as
composite \cite{comp}, GUT \cite{lqgut} and SUSY \cite{lqsusy} models, as
exotic particles which carry both color and lepton quantum numbers. In
general, they are either scalar or vector particles, with mass and coupling
constant to the standard fermions left as unknown parameters. Some
experimental contraints have been set on these parameters quite recently 
\cite{tevatron,CDF1,CDF2,lep,hera}.

Leptoquarks can be directly produced in $ep$ colliders but their pair
production at hadron colliders still has a clear advantage over any other
method: it is almost insensitive to the magnitude of the Yukawa coupling
which is unknown. Previous searches performed at the proton-antiproton
collider Tevatron (Fermilab) have excluded scalar leptoquarks with masses
below 175 GeV and 147 GeV for branching ratios of the leptoquarks to the
electron equal to 1 and 0.5 respectively \cite{tevatron}. For the second
generation, CDF sets limits and obtains 180 GeV (140 GeV) for $B=1$ $(0.5)$ 
\cite{CDF1}. Similarly, a limit of 99 GeV for $B=1$ was obtained by CDF for
third generation leptoquarks \cite{CDF2}. Some searches have also been
performed at LEP \cite{lep} and previous HERA runs have also contributed to
set limits \cite{hera}. A large machine like the LHC (with $\sqrt{s}=14$ TeV
and ${\cal L}=10$ fb$^{-1}$) should improve considerably such discovery
limits.

The cross sections for the pair production of scalar leptoquarks at hadron
colliders can be found in the literature \cite{sclq,dms,blum,kramer}.
However, a comprehensive study of the various QCD and electroweak
backgrounds which accompany leptoquark processes has been lacking up until
recently \cite{dms2}. Indeed, it is not trivial otherwise to estimate to
which extent the leptoquark signal will ``survive'' the QCD production of
heavy fermions, or jets produced along with the vector bosons $W$ and $Z$,
etc. Here, we shall consider events where both leptoquarks decay into an
electron plus quark, implying a 2 jets + $e^{+} e^{-}$ signature.

The purpose of this paper is to investigate the ability of the CERN Large
Hadron Collider (by using the design of the ATLAS and CMS experiments \cite
{atlas,cms}) to unravel the presence of scalar leptoquarks and to examine to
which extent the leptoquark signal can be distinguished from Standard Model
processes. We implement leptoquark data and related cross sections in the
ISAJET event generator and use the ISZRUN package contained in the Zebra
version to perform our selection cuts.

We consider the scalar leptoquarks contained in the supersymmetric grand
unified $E_6$ model (the low-energy limit of an $E_8\otimes E_8$ heterotic
string theory \cite{string}). In the $E_6$ model, each matter supermultiplet
lies in the fundamental ${\bf 27}$ representation, which contains, in
addition to the usual quarks and leptons (and their superpartners), new
particles such as two five-plets $(D,H)$ and $({\bar{D}},{\bar{H}})$ and an $%
SU(5)$ superfield singlet $N$. We focus on the superfields $D$ and ${\bar{D}}
$ which are two $SU(3)$ triplets and $SU(2)$ singlets with electric charges $%
-1/3$ and $+1/3$, respectively. Depending on the charge assignment chosen
for the superfields, $D$ and ${\bar{D}}$ can be taken to possess baryonic
number $\pm 1/3$ and leptonic number $\pm 1$. The scalar superpartners of
these superfields are the object of the present study. We thus consider
scalar leptoquarks with $Q=-1/3$. We restrict our study to the first
generation of fermions. The Yukawa interactions take the form: 
\begin{equation}
{\cal L}_Y=\lambda _L{\tilde{D}}^{c*}\left( e_Lu_L+\nu _Ld_L\right) +\lambda
_R{\tilde{D}}e_L^cu_L^c+\text{h.c.}  \label{L}
\end{equation}
where $c$ denotes the charge conjugate state, and ${\tilde{D}}$ is the
scalar superpartner of $D$. In this model, the $\lambda $ are independent
and arbitrary but we choose them to be equal to the electromagnetic charge,
following \cite{eeqq}. It is important to note that leptoquarks also
interact strongly. As we shall see, these interactions are mainly
responsible for their production in pairs.

In the following Section, we present the details of our simulation and the
selection cuts that we have chosen. Next, we elaborate on the expected
signature of the leptoquark signal and of the principal sources of
background: Drell-Yan and $t{\bar{t}}$ production. Finally, we summarize our
results and conclude in Section V.

\section{{\bf {Event simulation}}}

\subsection{Detector and calorimeter}

We use the toy calorimer simulation package ISZRUN contained in the Zebra
version of ISAJET \cite{isajet} to simulate the experimental conditions at
the LHC, with the ATLAS and CMS detectors in mind:

\begin{itemize}
\item  cell size: $\bigtriangleup \eta \times \bigtriangleup \phi
=0.05\times 0.05$,

\item  pseudorapidity range: $-5<\eta <5$,

\item  hadronic energy resolution: $50\%/\sqrt{E}\oplus 0.03$ for $-3<\eta
<3 $,\newline
\hspace*{49.5mm} $100\%/\sqrt{E}\oplus 0.07$ for $3<\mid \eta \mid <5$,

\item  electromagnetic energy resolution: $10\%/\sqrt{E}\oplus 0.01$ .%
\newline
\end{itemize}

\subsection{Kinematic cuts}

For the purposes of this work, hadronic showers are regarded as jets when
they

\begin{itemize}
\item  lie within a cone of radius $R=\sqrt{(\Delta \eta )^2+(\Delta \phi )^2%
}=0.7$,

\item  possess a transverse energy $E_T>25$ GeV,

\item  have a pseudorapidity $|\eta _j|\leq 3$.
\end{itemize}

Similarly, electrons are considered isolated if they

\begin{itemize}
\item  are separated from any jet by $R\geq 0.3$,

\item  have a transverse momentum $p_T>25$ GeV,

\item  have a pseudorapidity $|\eta _l|\leq 2.5$.
\end{itemize}

Our calculations are performed using the PDFLIB distribution functions of
Morfin and Tung (M-T B2) with $\Lambda =191$ MeV \cite{mt,pdflib}. The
choice of the distribution functions affects only slightly the cross
section. The calculations were repeated with more recent distribution
functions, namely CTEQ3M \cite{pdflib}. For leptoquark masses below 400 GeV,
the results remain practically unchanged; for higher masses, the cross
section is enhanced by at most 5\%.

\section{{\bf {Leptoquark Signal and backgrounds}}}

Here, we consider first-generation leptoquarks, which can decay into either
an $u$-quark and an electron, or into a $d$-quark and a $\nu _e$. For the
purposes of our calculations, we consider the case in which both occur with
equal probability ($B=0.5$). We also assume a Yukawa coupling of
electromagnetic strength $\alpha _Y=\alpha _{em}$ (in fact, it is a generic
feature of string-inspired models that the non-zero Yukawa coupling is of
the same order as the gauge coupling \cite{witten}). In fact, the Yukawa
coupling has only a very small impact on the pair production cross section.

\subsection{Leptoquark signal}

We analyze the pair production of scalar leptoquarks which arise from two
subprocesses: (1) quark-antiquark annihilation $(u_R+u_L^c\rightarrow {\ 
\tilde{D}}+{\tilde{D}}^{*}$ and $u_L+u_R^c\rightarrow {\tilde{D}}^{c*}+{%
\tilde{D}}^c)$, and (2) gluon fusion $(g+g\rightarrow {\tilde{D}}+{\ \tilde{D%
}}^{*}$ and $g+g\rightarrow {\tilde{D}}^{c*}+{\tilde{D}}^c)$ (see Fig. 1).
Whereas the first subprocess occurs in the $s$-channel (through the exchange
of a virtual gluon) and in the $t$-channel (virtual electron), subprocess
(2) arises via color gauge interactions from the trilinear term $gDD$ in the 
$s$-channel (through the exchange of a gluon) and in the $t$- and $u$%
-channels (exchange of virtual scalar leptoquarks), and from the quartic
term $ggDD$ in which two gluons annihilate to produce a pair of leptoquarks.

In our calculations, we omitted the soft-gluon correction $K$-factors \cite
{dms}, $K_{gg}=1+2\alpha _s\pi /3$ and $K_{q{\bar{q}}}=1-\alpha _s\pi /6,$
for gluon fusion and quark-antiquark annihilation respectively. Previous
studies suggest that the gluon fusion subprocess will dominate at the LHC
energies. Thus, we can expect a cross section enhancement factor ranging
from 1.22 to 1.19 for leptoquark masses of 200 GeV up to 1 TeV (assuming $%
\mu =M_{LQ}$ which is the choice of scale used throughout these
calculations). Recently, Kr\"{a}mer {\it et al.} \cite{kramer} have carried
out a complete NLO calculation of scalar leptoquark pair production; their
results are expressed in the form of an overall $K$-factor which essentially
reproduces the features of the soft-gluon $K$-factor approach with $\mu
=M_{LQ}$.

Leptoquark pair production can lead to three distinct signals:

\begin{enumerate}
\begin{enumerate}
\item  2 jets + $e^{+}e^{-}$,

\item  2 jets + ${\not{p}}_T$,

\item  2 jets + $e^{\pm }+{\not{p}}_T$.
\end{enumerate}
\end{enumerate}

The most striking of these signals is expected to be (a). In fact, signals
(b) and (c) are more cumbersome because many SM ($WW$, $WZ$, $ZZ$, $Zgg$ and 
$Zgq$ production) and SUSY processes have the same signatures. We therefore
restrict ourselves to 2 jets + $e^{+}e^{-}$. The background which comes
mainly from $t\bar{t}$ can be considerably reduced by requiring a cut on the
transverse energy of both the jets and the leptons. Here we shall impose the
same $E_T$ cut on the leptons and the jets.

\subsection{SM Backgrounds}

The most probable sources of background as identified by Refs. \cite{sclq}
are (1) $t\bar{t}$, (2) $Z^{*}jj$, (3) $ZZ$ and $WZ$ production. However,
our calculations have shown processes 2 and 3 (with an invariant mass cut on
the lepton pair: $81$ GeV $\leq M_{e^{+}e^{-}}\leq 101$ GeV) to be
negligible compared to (1). Therefore, we will restrict ourselves to $t\bar{t%
}$ (see Fig. \ref{feynbg}.) where the top is decaying into a $b$ quark, an
electron and a $\nu _e$. The presence of neutrinos implies in general a
missing transverse momentum ${\not{p}}_T$. In this case, it is natural to
expect the available transverse energy of the electron to be smaller on
average than that involved in the leptoquark process. Our calculations were
made using $M_t=175$ GeV.

\section{Leptoquark discovery reach at the CERN LHC}

Fig. \ref{ETPT200} shows the total cross section for 2 jets + $e^{+}e^{-}$
as a function of the mass for a transverse energy cut $E_T=200$ GeV. The
leptoquark signal (solid line) is plotted against the leptoquark mass
whereas the $t{\bar{t}}$ background (dashed line) is evaluated at $M_t=175$
GeV. The $5\sigma $ statistical significance is achieved for leptoquark
masses up to $750$ GeV. This limit also corresponds to $10$ leptoquark
events considering a luminosity of 10 fb$^{-1}$. Thus, we find a discovery
reach of $750$ GeV for leptoquarks that decay into electrons with $B=0.5$.
We can also evaluate the discovery limit for leptoquarks that decay with $%
B=1 $ by recalling that the cross section for the production of 2 jets + $%
e^{+}e^{-}$ is four times larger in this case. This leads to a discovery
reach of 1 TeV. Our discovery limits are somewhat lower than those recently
obtained in Ref. \cite{blum}. This is expected since the cuts that we have
applied to suppress the background have also reduced the signal cross
section.

One of the features of the leptoquark production process is the strong
correlation between the jet and the electron emerging from the same
leptoquark. In order to illustrate this fact, we look at the invariant mass
distribution of the lepton-jet pair ($M_{ej}$) of leptoquark pairs and the $t%
\bar{t}$ background. The reconstruction of leptoquarks from lepton-jet pairs
raise the problem of conveniently pairing each lepton with the right jet. A
method consist of associating the lowest-energy lepton with the
highest-energy jet, but it did not turn out to be the most efficient
procedure here. Instead, pairing the electrons and the jets using event
topology ({\it i.e.} matching an electron with its nearest-neighbor jet)
gave much better results. At the LHC, the pairs of leptoquarks in the mass
range under study ($M_{LQ}\ll \sqrt{s}$) are produced with very high kinetic
energy in opposite directions which explains why the decay products of each
leptoquark appear predominantly in opposite hemispheres. We present our
results for the invariant mass distribution of the lepton-jet pair ($M_{ej}$%
) in Figs. \ref{minvet200}-\ref{minvet100} for $E_T$ cuts of $100$ GeV and $%
200$ GeV respectively. The solid lines correspond to the leptoquark signal
with (a) $M_{LQ}=200$ GeV, (b) $M_{LQ}=500$ GeV and (c) $M_{LQ}=750$ GeV.
The dashed lines correspond to $t\bar{t}$ background. The lepton-jet
correlation is quite evident when looking at the peaks in the $M_{ej}$
distribution. In comparison, the background does not exhibit any such peaks
as can be expected from the presence of a missing ${\not{p}}_T$. The
signal-to-background ratio is optimal for an $E_T$ cut of $200$ GeV (Fig. 
\ref{minvet200}).

In order to emphasize the importance of the signal-to-background ratio near
the peak in the $M_{ej}$ distribution, we display in Fig. \ref{deltasigma}
the partial cross section integrated over a bin of width $\Delta M_{ej}=100$
GeV around $M_{ej}=M_{LQ}$ as a function of the invariant mass of the
electron-jet pair for $E_T=200$ GeV. The results are presented for a large
set of intermediate values of $M_{ej}=M_{LQ}$ within the range $100$ GeV $%
<M_{LQ}<1$ TeV. The leptoquark signal (solid line) exhibits a smooth
logarithmic behavior while the $t\bar{t}$ background shows some irregular
fluctuations around an approximatively constant value. Note that these
fluctuations can be misleading on a logarithmic plot as they turn out to be
rather small in magnitude. Comparing with Fig. \ref{ETPT200}, we find that
the signal-to-background ratio is increased by one order of magnitude in
Fig. \ref{deltasigma}. The $5\sigma $ statistical significance is achieved
for leptoquark masses up to 1 TeV.

In conclusion, we have presented the results of a complete analysis of the
first-generation scalar leptoquark pair production within the context of an $%
E_6$ model. We have also calculated the importance of the various Standard
Model backgrounds which have the same signature. The leptoquark signal was
found to be dominant over the $t\bar{t}$ background for leptoquark masses up
to $750$ GeV. We have evaluated our leptoquark discovery limit for the
optimal case $E_T=200$ GeV. We found a leptoquark discovery reach of $750$
GeV ($1$ TeV) for a branching ratio of $B(LQ\rightarrow eq)=0.5$ ($B=1$).

\acknowledgements
We are indebted to Pr. J. Pinfold for useful comments. The authors would
like to thank Pr. Frank Paige for answering numerous questions during the
installation of ISAJET. This research was supported by the Central Research
Fund of the University of Alberta, contract no CRF-GEN 81-53428, the Natural
Sciences and Engineering Research Council of Canada, and by the Fonds pour
la Formation de Chercheurs et l'Aide \`{a} la Recherche du Qu\'{e}bec.

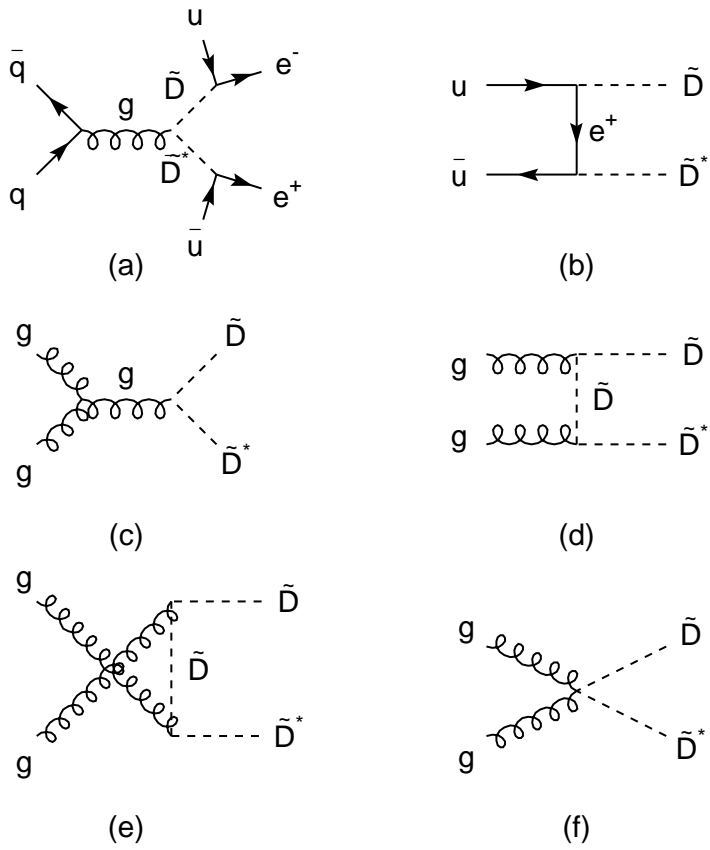
\begin{figure}[tbp]
\caption{Feynman diagrams for leptoquark pair production via ((a), (b)) $q 
\bar{q}$ annihilation and ((c), (d), (e), (f)) gluon fusion.}
\label{feynlq}
\end{figure}

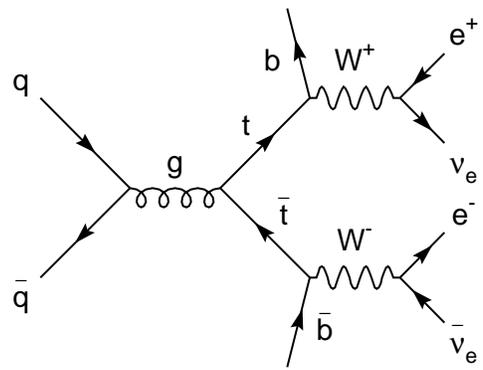
\begin{figure}[tbp]
\caption{Feynman diagram for $t \bar{t}$ production.}
\label{feynbg}
\end{figure}

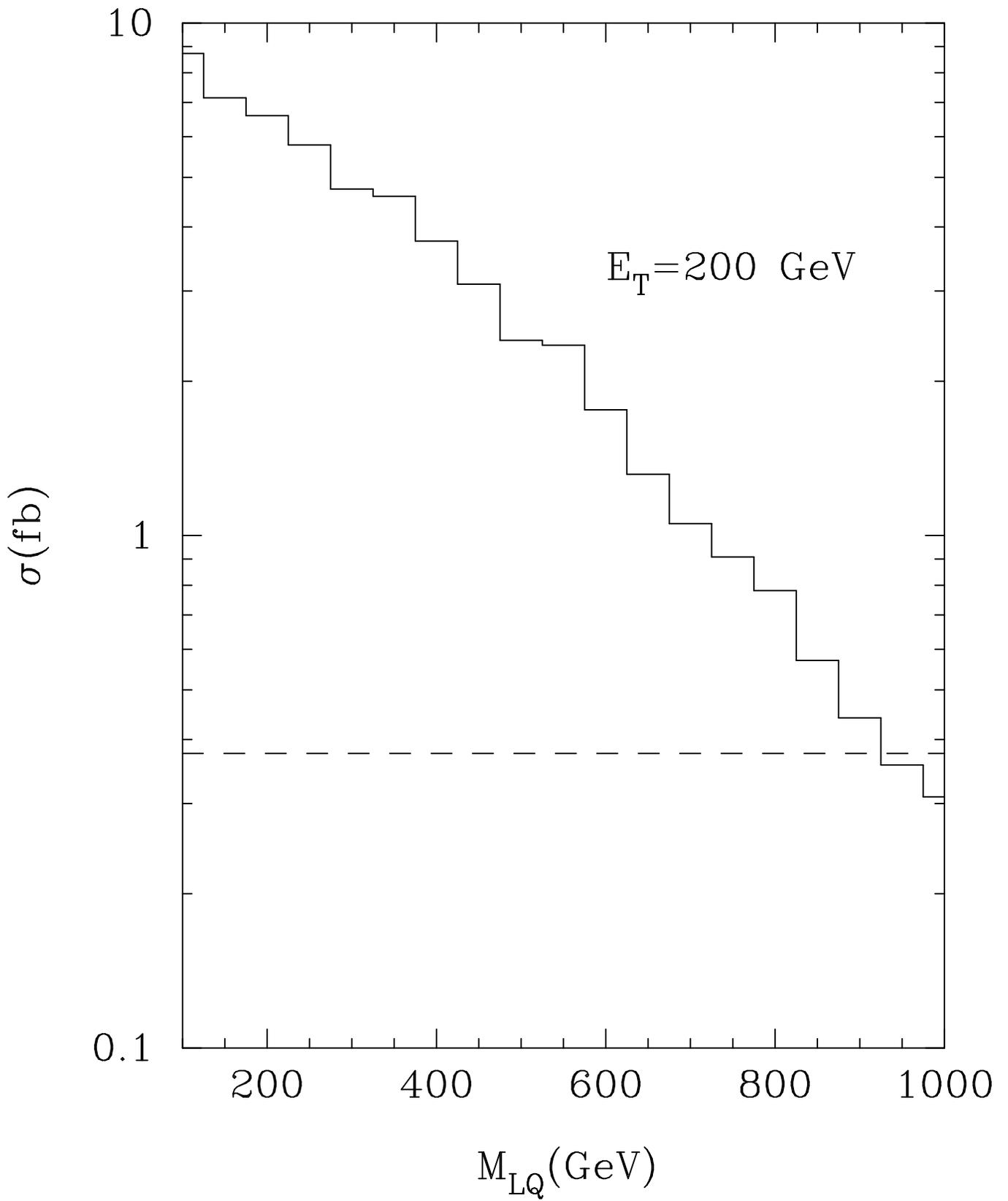
\begin{figure}[tbp]
\caption{Integrated cross section for the production of 2 jets + $e^{+}
e^{-} $ as a function of the leptoquark mass for $E_{T}=200$ GeV. The full
line corresponds to the leptoquark signal versus the leptoquark mass ($%
M_{LQ} $) and the dashed line to $t{\bar t}$ background for $M_{t}$=175 GeV.}
\label{ETPT200}
\end{figure}

\begin{figure}[tbp]
\caption{Distribution of the invariant mass of the lepton-jet pair ($M_{ej}$%
) for the production of 2 jets + $e^{+} e^{-}$ for $E_{T}=100$ GeV. The
solid lines correspond to the leptoquark signal with (a) $M_{LQ}$=200 GeV,
(b) $M_{LQ}$=500 GeV and (c) $M_{LQ}$=750 GeV. The dashed lines correspond
to $t{\bar t}$ background.}
\label{minvet100}
\end{figure}

\begin{figure}[tbp]
\caption{Same as figure 4 but for $E_{T}=$200 GeV.}
\label{minvet200}
\end{figure}

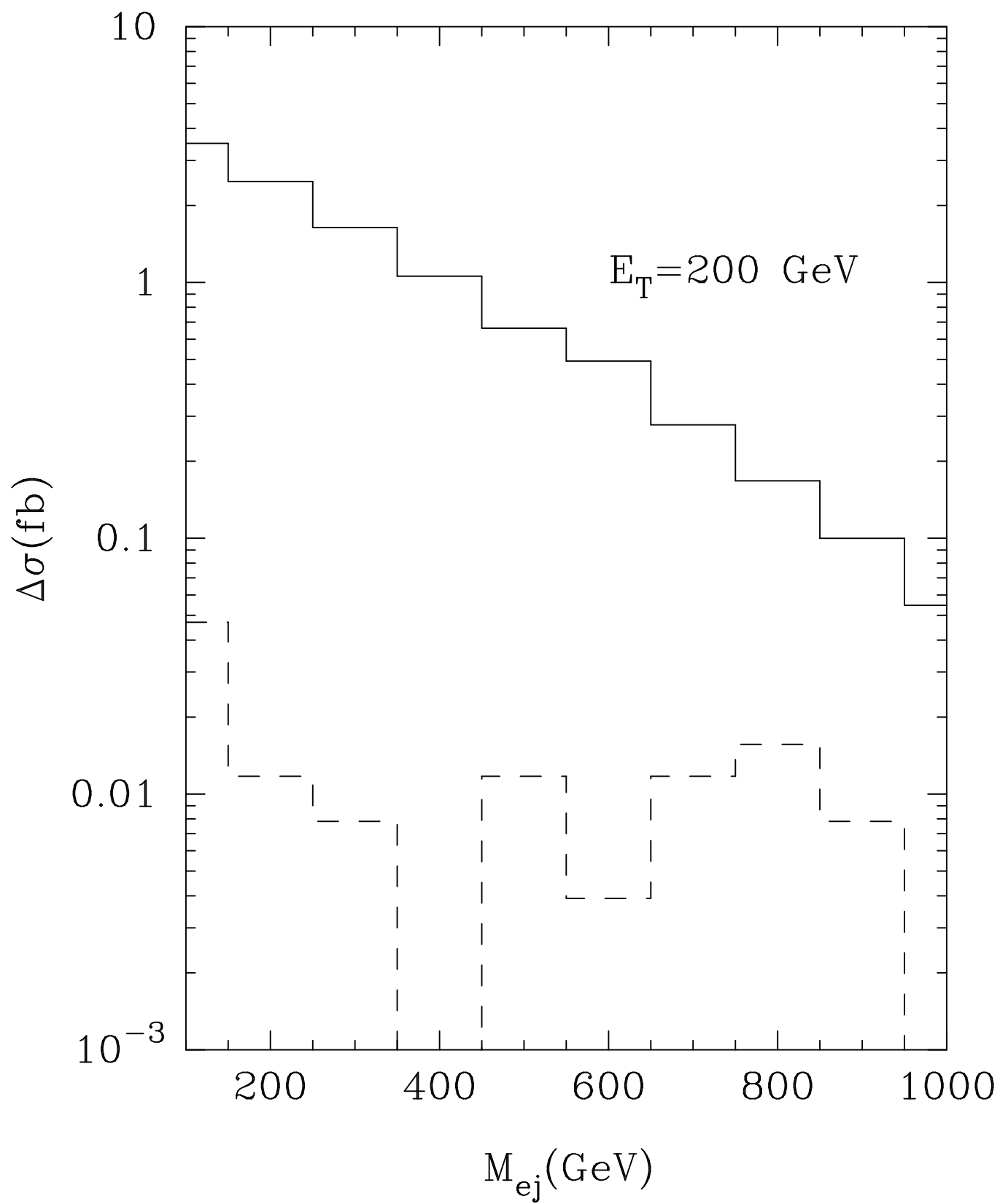
\begin{figure}[tbp]
\caption{Partial cross section within a bin of width $\Delta M_{ej}$=100 GeV
around $M_{ej}=M_{LQ}$ as a function of the invariant mass of the
electron-jet pair for $E_{T}=200$ GeV. The full line corresponds to the
leptoquark signal and the dashed line to $t{\bar t}$ background for $M_{t}$%
=175 GeV.}
\label{deltasigma}
\end{figure}

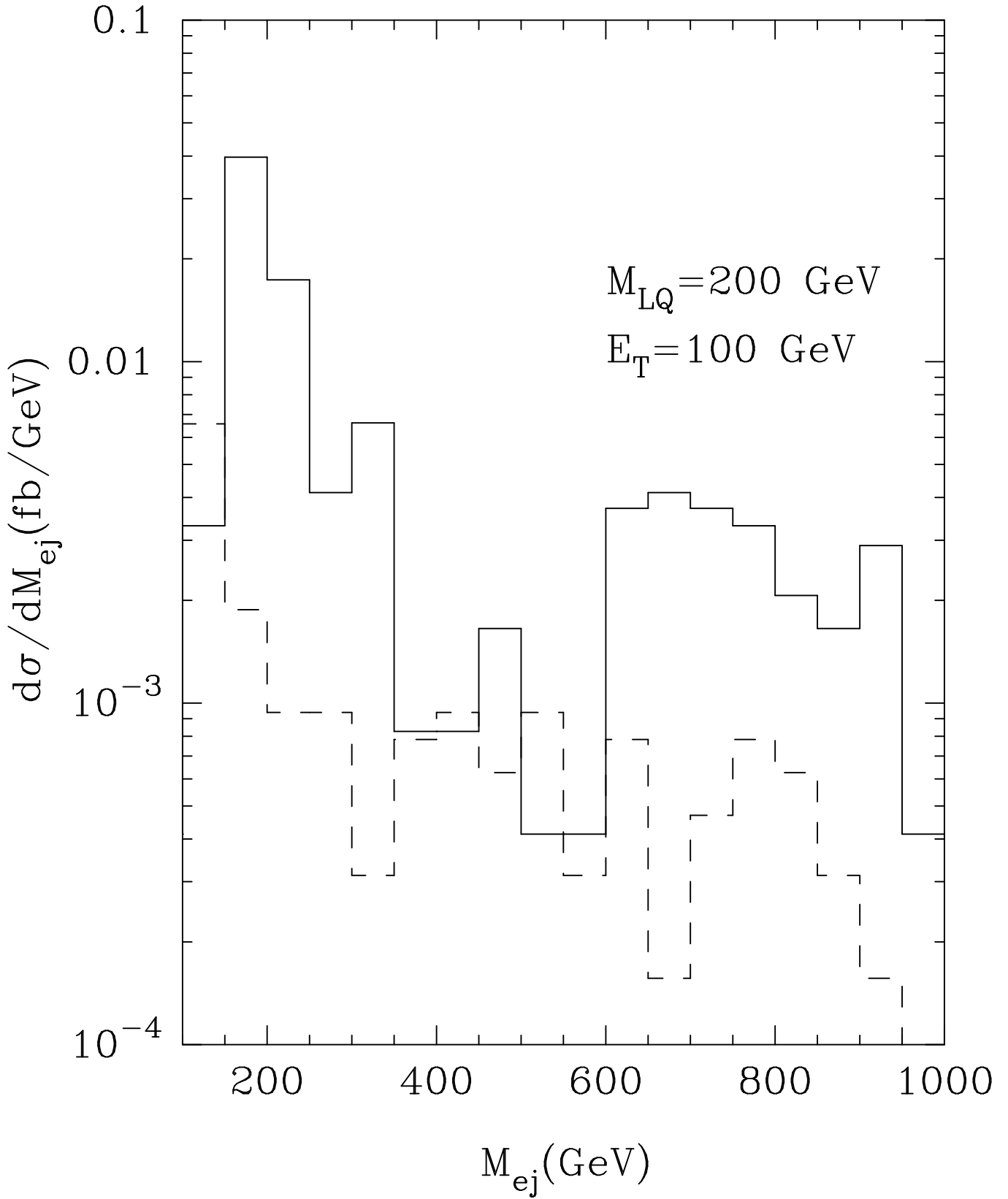
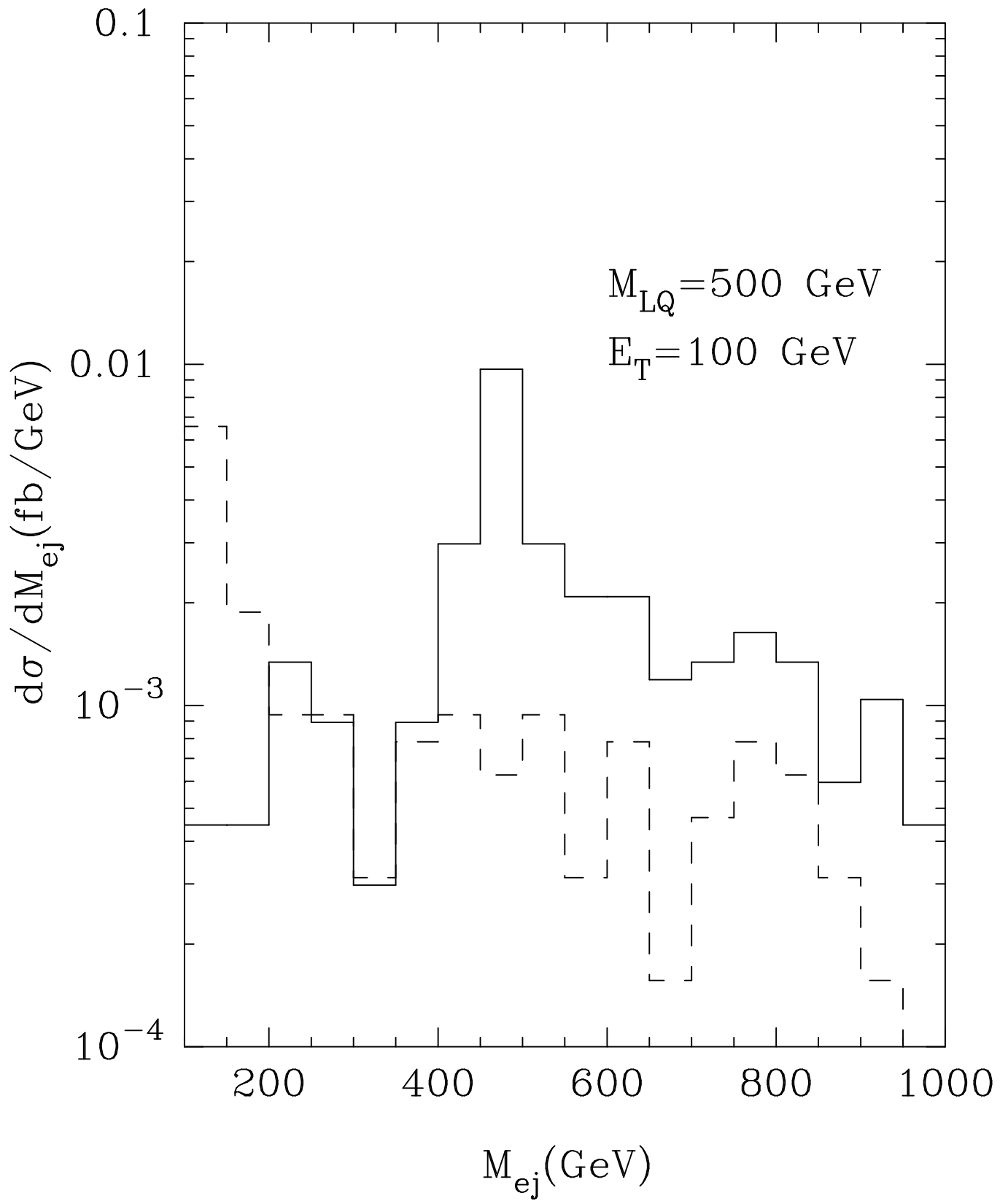
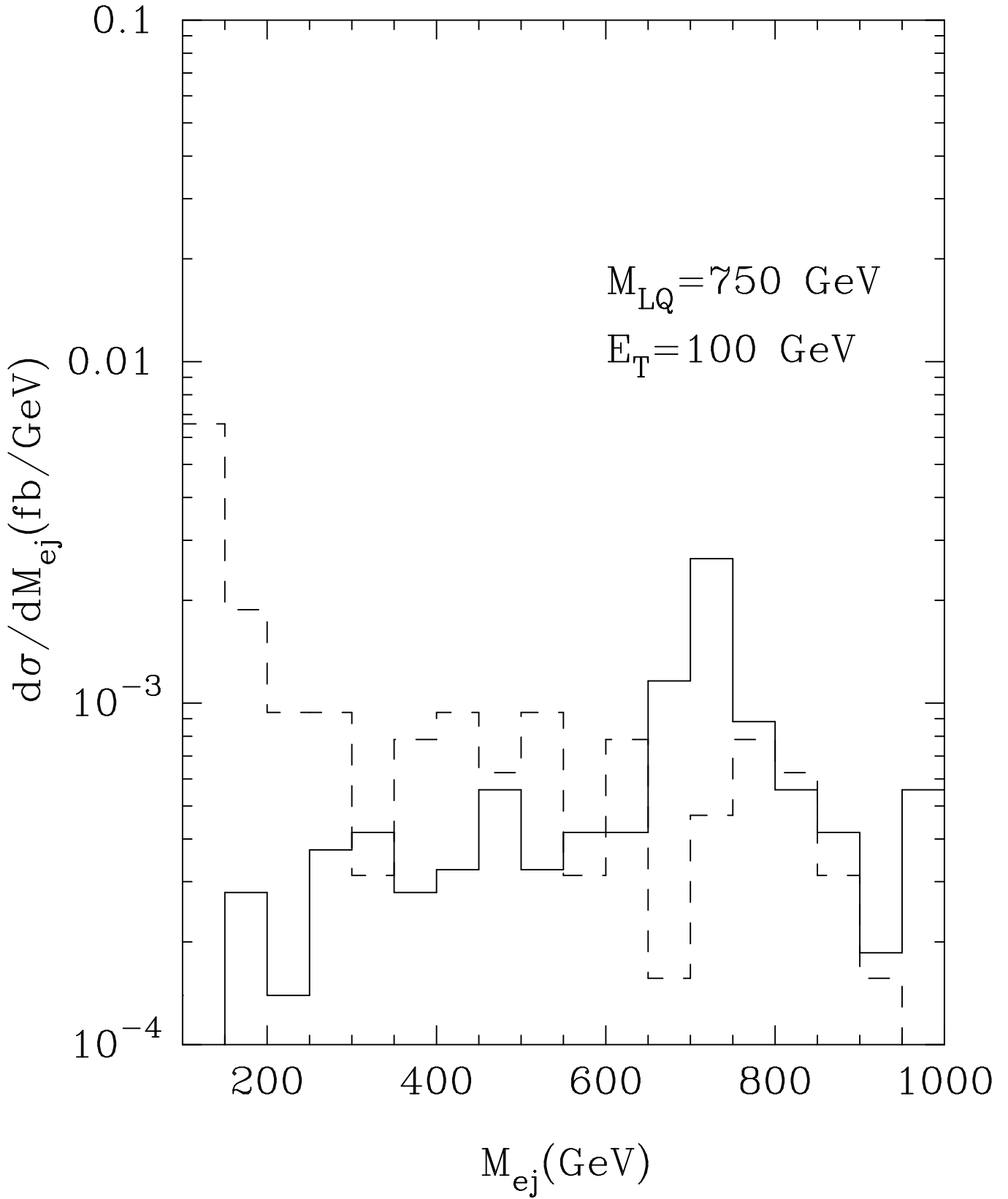
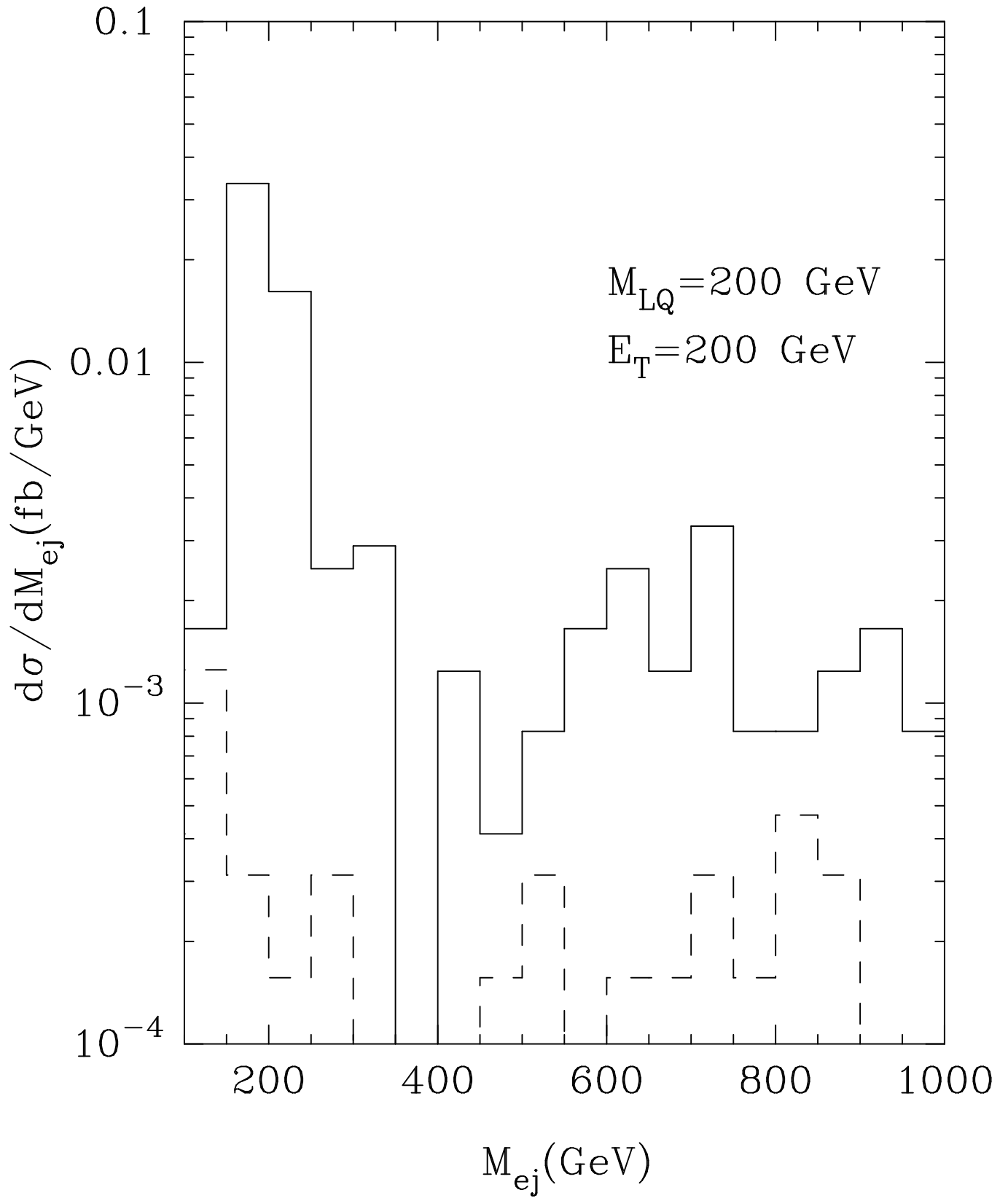
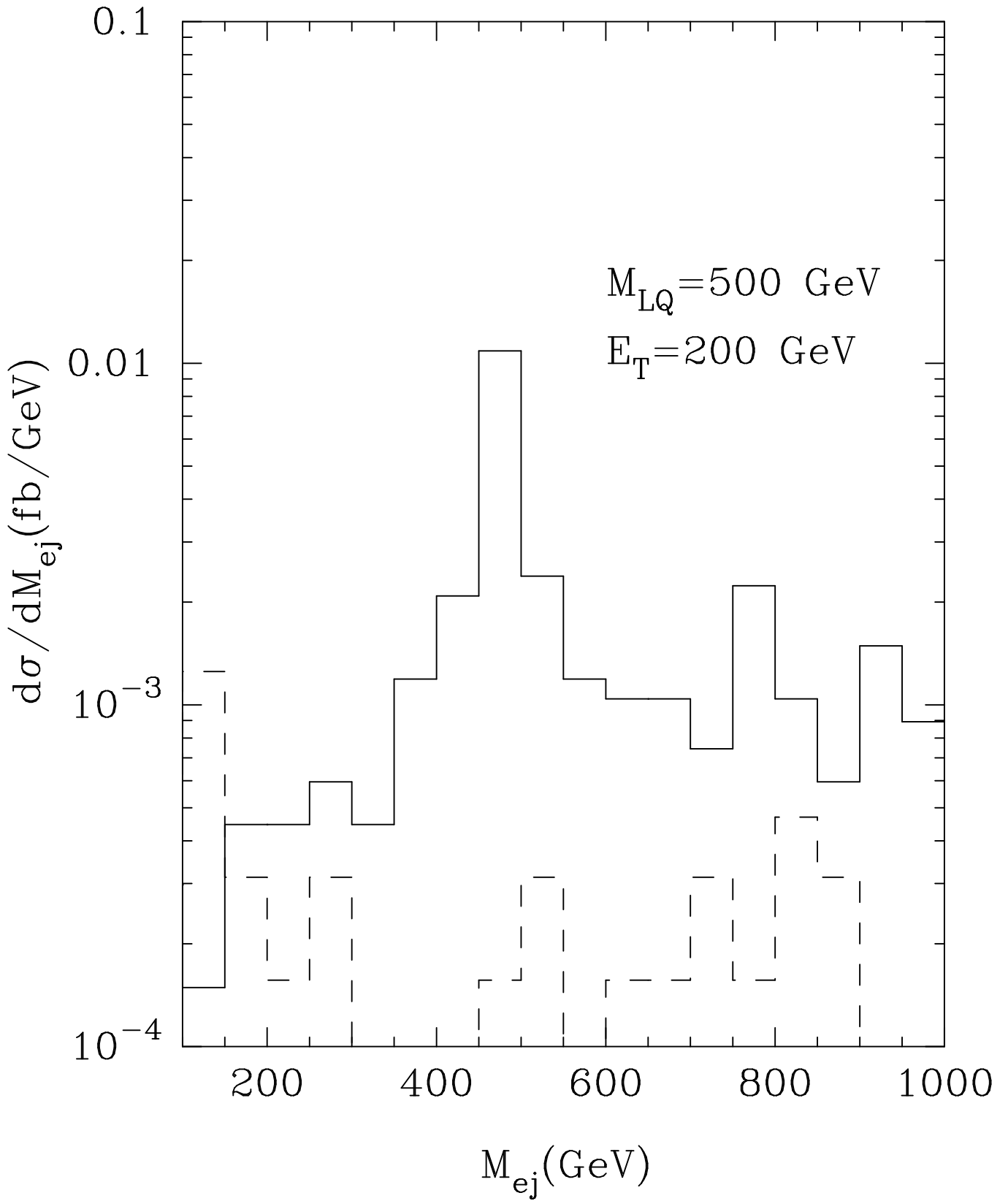
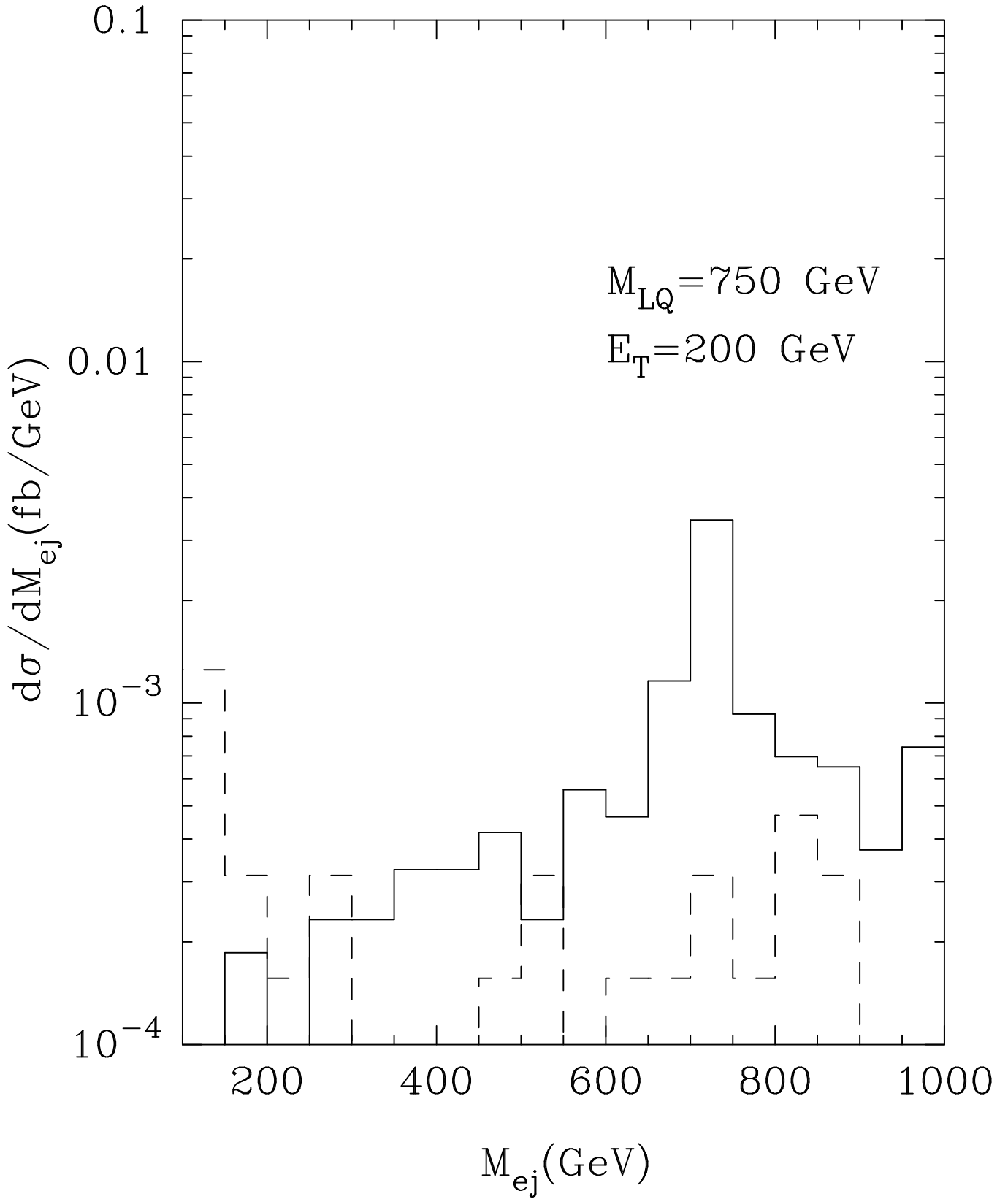
\end{document}